\documentstyle[12pt]{article}
\begin{document}

\title{New Analysis of Cumulant Moments in $e^+e^-$ Collisions 
by SLD Collaboration by Truncated Multiplicity Distributions}
\author{Naomichi Suzuki, Minoru Biyajima$^{1}$ and Noriaki 
Nakajima$^{1}$ \\
 ${}^{}$Matsusho Gakuen Junior College, Matsumoto 390-12, 
Japan  \\  
 ${}^{1}$Department of Physics, Shinshu University, 
Matsumoto 390, Japan }
\date{}
\maketitle
 
\begin{abstract}
Newly reported normalized cumulant moments of charged particles 
in $e^+e^-$ collisions by the SLD collaboration are analyzed by 
the truncated modified negative binomial distribution (MNBD) and 
the negative binomial distribution (NBD).  Calculated result by 
the MNBD describes the oscillatory behavior of the data  much 
better than that by the NBD.
\end{abstract}

PACS number(s): 13.65.+i, 34.10.+x
 
\vspace{5mm}
Normalized cumulant moments of negatively charged particles and 
charged particles obtained from the data in $e^+e^-$ collisions 
were analyzed by the truncated modified negative binomial 
distribution (MNBD) in \cite{suzu96}.  Those moments show 
oscillatory behaviors as the rank of the moment increases, and 
are well described by the MNBD.
 
The preliminary data of cumulant moments normalized by factorial 
moments, $H_j$ $\,(j=1,2,\cdots)$  of charged particles reported 
by the SLD Collaboration at $\sqrt{s}=91$ GeV were also analyzed 
in \cite{suzu96}.  The SLD did not explicitly show us the 
multiplicity distribution. Therefore, we calculated $H_j$ moments 
using the observed values of the average charged multiplicity, 
$C_2$ moment, and the maximum of the negatively charged particles 
$n_{\rm max}=27$.  The data were only qualitatively explained 
by our calculation.  
 
Recently, the new data for the $H_j$ moments are reported by the SLD  
\cite{abe96} and \cite{priv96}. In this brief report, 
they are re-analyzed by the MNBD.
 
The MNBD is given by 
 \begin{eqnarray}
    P(0) &=& \frac{\left(1+r_1 \right)^N}{\left(1+r_2 \right)^{N}},
                                \nonumber \\
    P(n) &=& \frac{1}{n!} \left( \frac{r_1}{r_2} \right)^N
              \sum_{j=1}^N {}_N C_j \frac{\Gamma(n+j)}{\Gamma(j)}
              \left(\frac{r_2-r_1}{r_1} \right)^j
      \frac{ r_2^n }{ \left( 1+r_2 \right)^{n+j} },
                \quad  n=1,2, \cdots,    \label{eq:kq1}
 \end{eqnarray}
where $N$ is a positive integer, $r_1$ is real, 
and $r_2>0$.  
 
At first, Eq.(\ref{eq:kq1}) is applied to the multiplicity 
distribution of negatively charged particles.
 In order to calculate cumulant moments, factorial moments of 
charged particles are calculated as
 \begin{eqnarray}
   f_j^{{\rm ch}} &=& <n_{\rm ch}(n_{\rm ch}-1)\cdots 
(n_{\rm ch}-j+1)>
                           \nonumber  \\
       &=& \sum_{n}^{n_{\rm max}}\, 2n(2n-1)\cdots(2n-j+1)\,P(n),
              \quad j=1,2,\cdots,   \label{eq:kq2}
 \end{eqnarray}
where $n_{\rm max}$ denotes the maximum of the observed negatively 
charged multiplicity. 
 
The j-th order normalized cumulant $K_j$ of charged particles is 
expressed by the normalized factorial moments $F_l \quad(l=1,2,
\cdots)$ of the charged particles as;
 \begin{eqnarray}
        K_1 &=& F_1,      \nonumber  \\
        K_j &=& F_j
        + \sum_{m=1}^{j-1} {}_{j-1}C_{m-1}\,F_{j-m}\,K_m,
                  \qquad j=2,3,\cdots,    \label{eq:kq3}
 \end{eqnarray}
where
 \begin{eqnarray*}
    F_j = \frac{f_j^{\rm ch}}{<n_{\rm ch}>^j}.
 \end{eqnarray*}
The $H_j$ moment is defined by
 \begin{eqnarray}
    H_j = K_j/F_j.   \label{eq:kq4}
 \end{eqnarray}

 The parameters in Eq.(\ref{eq:kq1}) are determined by 
the minimum chi-square ($\chi^2_{\rm min}$) fit to the observed 
multiplicity distribution \cite{priv96} of negatively charged 
particles with 
$n_{\rm max}=25$.  The result is shown in Table 1.
 
The $H_j$ moment is calculated from Eqs.(\ref{eq:kq1}), 
(\ref{eq:kq2}), (\ref{eq:kq3}) and (\ref{eq:kq4}). The result is 
shown in Fig.1.  Calculation with the MNBD denoted by the solid 
line well reproduces the oscillatory behavior of the data.
 
 For comparison, we try to fit the new observed multiplicity 
distribution using the negative binomial distribution (NBD);
 \begin{eqnarray}
    P(n) = \frac{\Gamma{(k+n)}}{\Gamma{(k)}\Gamma{(n+1)}} 
              \left( \frac{<n>}{k} \right)^n
              \left( 1+\frac{<n>}{k} \right)^{-n-k},
                \quad  n=0,1,2, \cdots.    \label{eq:kq5}
 \end{eqnarray}
However, the reasonable $\chi^2_{\rm min}$ value cannot be found 
in the region with $k>0$ \footnote{ The data are analyzed by a 
weighted superposition of two NBD's \cite{giov96} }.  See Table 1.
 
 We also analyze the data applying Eq.(\ref{eq:kq1}) to the 
charged particles; only the even terms in Eq.(\ref{eq:kq1}) are 
used with a normalization factor $C$ \cite{ugoc95};
 \begin{eqnarray}
   f_j^{{\rm ch}} = \sum_{n}^{n_{\rm max}}\, 2n(2n-1)
\cdots(2n-j+1)\,C\,P(2n),
         \quad j=1,2,\cdots,   \label{eq:kq6}
 \end{eqnarray}
The factor $C$ is determined by the following equation,
 \[   C \sum_{n}^{n_{\rm max}} P(2n) = 1.      \]
Then, the $H_j$ moment is calculated from Eqs.(\ref{eq:kq6}), 
(\ref{eq:kq3}) and (\ref{eq:kq4}).
 
 The parameters determined by the $\chi^2_{\rm min}$ fit to the 
multiplicity distribution are also shown in Table 1. The results 
are depicted in Fig.2. The result obtained from the MNBD, 
expressed by the solid line, well describes the behavior of 
the data. However, the result from the NBD, expressed by the 
dashed line, cannot explain the data.
 
\vspace{5mm}
{\bf Acknowledgements}
 
The authors thank Jingchen  Zhou and H. Masuda for their kind 
correspondences.   M. B. is partially supported by the 
Grant-in Aid for Scientific Research from the Ministry of 
Education, Science and Culture (No. 06640383).  N. S. thanks 
for the financial support by Matsusho Gakuen Junior College. 
%
 
%

%
 
\newpage
 
\vspace{1cm}
\begin{flushleft}
{\large Table caption}
\end{flushleft}
\begin{itemize}
 
\item[Table 1] The parameters of the MNBD and the NBD 
used in the analysis of the cumulant moments. The sign 
"$-$" and "$\pm$" denote negatively charged particles and 
charged particles, respectively.  

\end{itemize}
\begin{flushleft}
{\large Figure captions}
\end{flushleft}
\begin{itemize}
 
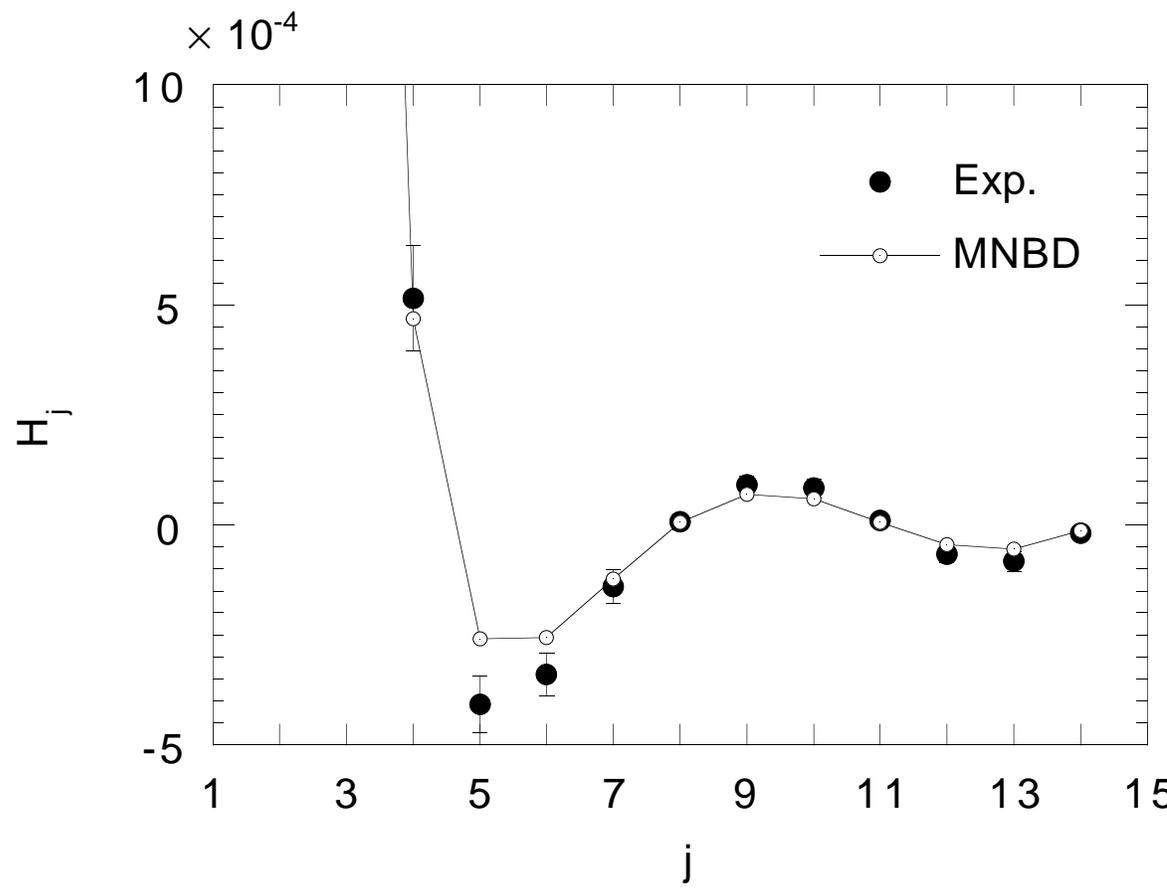
\item[Fig. 1] The normalized cumulant moments $H_j$ $\,
(j=1,2,\cdots)$ of charged particles in $e^+e^-$ 
collisions\cite{priv96}.  
 The full circles are obtained from the data.  The solid line is 
obtained from the MNBD, which is applied to the negatively charged 
multiplicity distribution. The calculation based on the NBD is 
not shown here, because $\chi^2_{\rm min}$ value is fairly 
large and $k<0$.

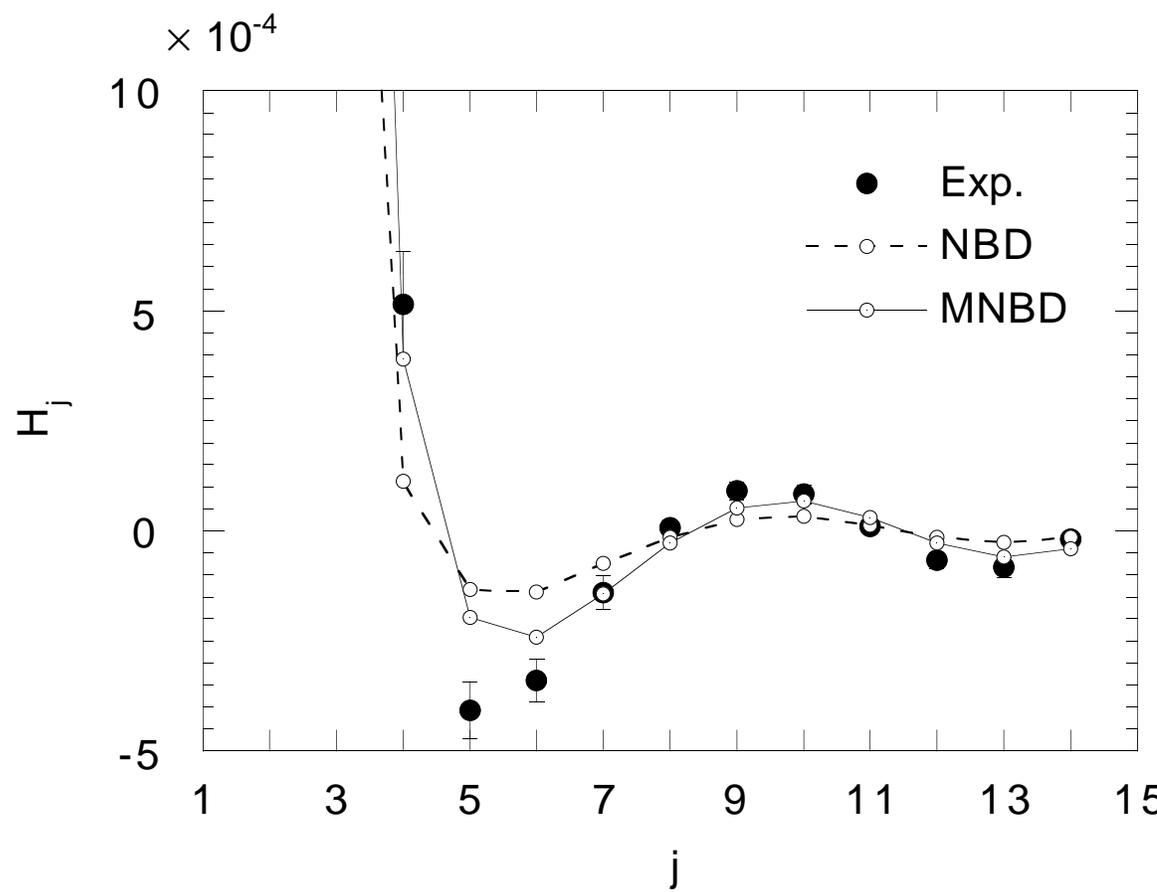
\item[Fig. 2] The normalized cumulant moments $H_j$ $\,
(j=1,2,\cdots)$
of charged particles in $e^+e^-$ collisions\cite{priv96}.
 The full circles are obtained from the data.  The solid 
line is obtained from
the MNBD, and the dashed line from the NBD. They are applied 
to the charged multiplicity distribution.  The value of 
$\chi^2_{\rm min} = $ 119.3  is attributed to the latest data
\cite{priv96}.
\end{itemize}
 
\newpage
\vspace{2cm}
\begin{center}
 \begin{tabular}{lccccc} \hline
  MNBD & $N$     & $r_1$   & $r_2$  &  $\chi^2_{\rm min}$  & {} \\
   \hline
     {} &  { 8}  & {-0.6873}$\pm$ 0.0027  &
             { 0.6162}$\pm$ 0.0029  & 54.1 & $-$   \\
     {} &  {13}  & -0.3580$\pm$ 0.0054 & 
              {1.249}$\pm$ 0.006 & 26.8 &  $\pm$   \\
          \hline  
          \hline
     NBD & {} &  $<n>$   &   $k$     &  $\chi^2_{\rm min}$ & {}   \\
          \hline
     {}   & {} & 10.64 $\pm$ 0.02  & -70.00 $\pm$ 2.57   &  {1066  }
                         & $-$     \\
     {}  & {} & 21.01 $\pm$ 0.03  &  24.43 $\pm$ 0.30   &  { 119.3}
                         & $\pm$   \\
          \hline
 \end{tabular}

\vspace{5mm}
  Table 1
\end{center}
\end{document}